# Properties of Periodic Fibonacci-like Sequences

Alexander V. Evako
Email address: evakoa@mail.ru

**Abstract**
Generalized Fibonacci-like sequences appear in finite difference approximations of the Partial Differential Equations based upon replacing partial differential equations by finite difference equations. This paper studies properties of the generalized Fibonacci-like sequence $F_{n+2} = A + BF_{n+1} + CF_n$. It is shown that this sequence is periodic with the period $T \in (2, \infty)$ if $C = -1$, $|B| < 2$.

**Keywords:** Generalized Fibonacci-like sequence, Periodicity, Discrete Dynamical System, PDE, Graph, Network, Digital Space

## 1. Preliminaries

Generalized Fibonacci sequences are widely used in many areas of science including applied physics, chemistry and biology. Over the past few decades, there has been a growth of interest in studying properties periodic Fibonacci sequences modulo m, Pell sequence and Lucas sequence. Many interesting results have been received in this area by researchers (see, e.g., [1, 4, 8, 10, 14]). Fibonacci like sequences are basic elements in computational solutions of partial differential equations using finite difference approximations of the PDE [6, 11], and in partial differential equations on networks and graphs [2, 7, 9]. The goals for this study is to investigate properties of the Fibonacci-like sequences, which are structural elements in discrete dynamic processes that can be encoded into graphs. The property we are interested in this study is the periodicity of a the Fibonacci-like sequence.

In Section 2, we study a periodic continuous function $f(x)$, where $f(x+1) = A + Bf(x) + Cf(x-1)$. We show that if $f(x)$ is periodic with period $T > 2$ then $|B| < 2$, $C = -1$. We apply the obtained results to the sequence, which is defined by recurrence relation $F_{n+2} = A + BF_{n+1} + CF_n$ with $F_0 = a$, $F_1 = b$, $n \geq 0$, where A, B C, a and b are real numbers. We show that if $|B| < 2$, $C = -1$ then $\{F_n\}$ is periodic with the period $T > 2$. Then we present numerical examples of periodic Fibonacci-like sequences.

## 2. Periodic Fibonacci-like continuous functions and sequences

Consider properties of periodic continuous function f(x) on domain R provided that $f(x+1) = A + Bf(x) + Cf(x-1)$.

**Proposition 1**

Let f(x) be a continuous function on domain R, and let
$$f(x+1) = A + Bf(x) + Cf(x-1) \qquad (1)$$
for any $x \in R$. If f(x) is a periodic function with period $T > 2$ then $|B| < 2$, $C = -1$, $T = \frac{2\pi}{\omega}$, $\omega = \arccos\frac{B}{2}$.

Proof.
Let f(x) is periodic with period $T = 2\pi/\omega$. Consider the Fourier series representing f(x).
$$f(x) = a_0 + \sum_{n=1}^{\infty} a_n \cos\omega nx + b_n \sin\omega nx \qquad (2)$$
Using (2), it is easy to see that
$f(x+1) - A - Bf(x) - Cf(x-1) = c_0 + \sum_{n=1}^{\infty} c_n \cos\omega nx + d_n \sin\omega nx$, where
$c_0 = a_0 - Ca_0 - Ba_0 - A \qquad (3)$
$c_n = a_n(\cos\omega n - C\cos\omega n - B) + b_n(1+C)\sin\omega n \qquad (4)$

$$d_n = -a_n(1+C)\sin\omega n + b_n(\cos\omega n - C\cos\omega n - B), \qquad (5)$$

Since $f(x+1) - A - Bf(x) - Cf(x-1) = 0 \; for \; \forall x$ then for $\forall n$, coefficients $c_0$, $c_n$ and $d_n$ are equal to zero.

$$c_0 = c_n = d_n \equiv 0. \qquad (6)$$

Let n=1.

$$c_1 = a_1(\cos\omega - C\cos\omega - B) + b_1(1+C)\sin\omega = 0$$
$$d_1 = -a_1(1+C)\sin\omega + b_1(\cos\omega - C\cos\omega - B) = 0.$$

Since f(x) is periodic with basic period $T = \frac{2\pi}{\omega}$ then $a_1 \neq 0 \; and/or \; b_1 \neq 0$, which means that

$$\begin{cases} \cos\omega - C\cos\omega - B = 0 \\ (1+C)\sin\omega = 0 \end{cases} \qquad (7)$$

Suppose that $\sin\omega \neq 0$, i.e., $\omega \neq k\pi$, $T \neq \frac{2}{k}$. It follows from (7) that $1 + C = 0$, $\cos\omega = \frac{B}{1-C}$, $|\cos\omega| < 1$. Therefore, $C = -1$, $B = 2\cos\omega$, $|B| < 2$. Hence, $\omega = \arccos\frac{B}{2}$ and $T = \frac{2\pi}{\omega}$.

Let n>1.

$$c_n = a_n(\cos\omega n - C\cos\omega n - B) + b_n(1+C)\sin\omega n = 0 \qquad (8)$$
$$d_n = -a_n(1+C)\sin\omega n + b_n(\cos\omega n - C\cos\omega n - B) = 0$$

For the same reason as above, if $\cos\omega n - C\cos\omega n - B \neq 0$ then $a_n = b_n \equiv 0$, if $\cos\omega n - C\cos\omega n - B = 0$ then $a_n \not\equiv 0 \; and/or \; b_n \not\equiv 0$. Note that $\cos\omega n - C\cos\omega n - B = 0 \; if \; (n \pm 1)\omega = 2\pi k$, $n, k \in N$.

Finally, find $a_0$ from equation (3). Notice that $C = -1$, $|B| < 2$.

$$a_0 = \frac{A}{1-C-B} = \frac{A}{2-B}$$

Thus, if $|B| < 2 \; and \; C = 1$ then f(x) is periodic with basic period $T = \frac{2\pi}{\omega}$ and $\omega = \arccos\frac{B}{2}$.

$f(x) = a_0 + a_1\cos\omega x + b_1\sin\omega x + \sum_{n\in G} a_n\cos\omega nx + b_n\sin\omega nx$, where $n \in G$ if $(n \pm 1)\omega = 2\pi k$, $n, k \in N$. This completes the proof. □

It is easy to check directly the following result.

**Proposition 2**

Let f(x) be a continuous periodic function on domain R with period $T = \frac{2}{k}$, $k \in N, k \neq 0$, and for any $x \in R$, $f(x+1) = A + Bf(x) + Cf(x-1)$.
Then
$f(x+1) = f(x)$ if $k = 2n$.
$f(x+1) = D - f(x)$ if $k = 2n+1$.

**Remark 1**

If f(x) is a continuous periodic function on domain R with period $T \leq 2$, $T \neq \frac{2}{k}$, $k \in N, k \neq 0$, then $f(x+1) \neq A + Bf(x) + Cf(x-1)$, $x \in R$.

We now outline a correspondence between the number B and the period T of function $f(x+1) = A + Bf(x) - f(x-1)$, where $|B| < 2$. Let's investigate this question graphically. The graph $T = T(B) = 2\pi \left(\arccos\frac{B}{2}\right)^{-1}$ is shown in fig. 1. Obviously, $\lim_{B\to 2} T = \infty$, $\lim_{B\to -2} T = 2$.

In a numerical solution of the one-dimensional wave equation by the finite difference method, the spatial domain and the temporal domain are replaced by a set of mesh points, the second-order derivatives can be replaced by central differences [2, 6, 7, 9]. The solution of the wave equation is replaced by the mesh function, which approximates the exact solution at the mesh points. The obtained discretization scheme has the form $u_i^{n+1} = 2u_i^n + c^2(u_{i-1}^n - 2u_i^n + u_i^n) - u_i^{n-1}$, where $c = k(\Delta t/\Delta x)$. By construction, this is a periodic in time sequence of real numbers..
Periodic generalized Fibonacci sequences modulo m have been studied in a number of papers (see e.g. [3, 5, 12, 13]). Let us consider the following sequence
$F_{n+2} = A + BF_{n+1} + CF_n$, $n \geq 0$, $F_0 = a$, $F_1 = b$ \qquad (11)
In sequence (11), $A, B, C, a$ and $b$ are real numbers. In particular, $A, B, C, a$ and $b$ can be integer. The following propositions are a direct consequence of propositions 1 and 2.

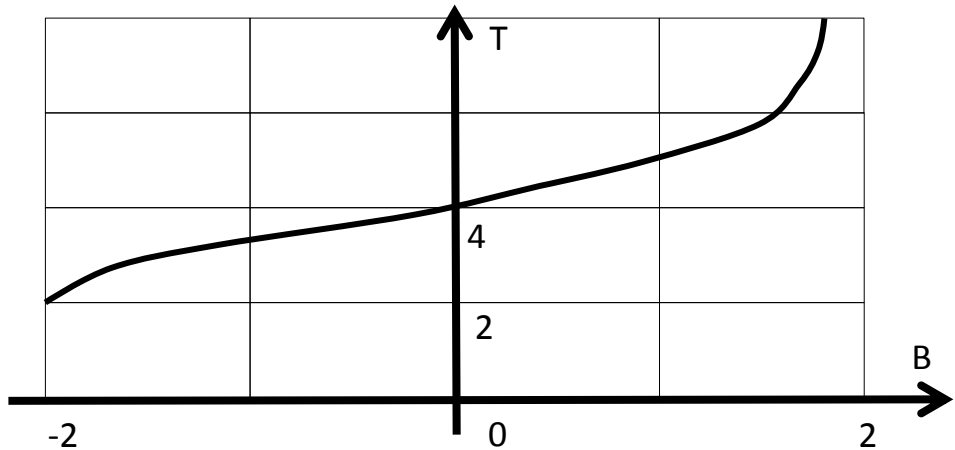

Figure 1. The period profile T of function f(x) and sequence $F_n$, $|B|<2$.

**Proposition 3**
  Let $F_{n+2} = A + BF_{n+1} + CF_n$, $n \geq 0$, $F_0 = a$, $F_1 = b$, be a sequence of real numbers,
  $(A, B, C, a, b) \in R$. If $|B| < 2$, $C = -1$ then $\{F_n\}$ is a periodic sequence with period $T = \frac{2\pi}{\omega}$, $\omega = arccos\frac{B}{2}$. $T > 2$.
Proof.
Consider the function $f(x + 1) = A + Bf(x) + Cf(x - 1)$ where $f(0) = a$, $f(1) = b$. Then $f(n + 2) = F_{n+2}$, $f(n + 1) = F_{n+1}$, $f(n) = F_n$ for $n \geq 0$.
According to proposition 1, if $|B| < 2$, $C = -1$, then f(x) is periodic with period $T = \frac{2\pi}{\omega}$, $\omega = arccos\frac{B}{2}$. $T > 1$, $T \neq 2$. Therefore, sequence $\{F_n\}$ is also periodic with period $T = \frac{2\pi}{\omega}$.
This completes the proof. □
Consider a special case when the period T=1, 2. The following proposition is a direct consequence of proposition 2.

**Proposition 4**
  Let $F_{n+2} = A + BF_{n+1} + CF_n$, $n \geq 0$, $F_0 = a$, $F_1 = b$, be a sequence of real numbers, $(A, B, C, a, b) \in R$.
  If $A = 0$, $B = 1$, $C = 0$ then $\{F_n\}$ is a periodic sequence with period $T = 1$, $F_{n+1} = F_n$, $n \geq 0$, $F_0 = F_1 = a$.
  If $C \neq 1$, $B = C - 1$ then $\{F_n\}$ is a periodic sequence with period $T = 2$,

$$F_{n+1} = D - F_n, \quad D = \frac{A}{1-C}, \quad n \geq 0, \quad F_0 = a, \quad F_1 = D - a.$$

Consider examples of the periodic generalized Fibonacci sequence $F_{n+2} = A + BF_{n+1} + CF_n$, $n \geq 0$, $F_0 = a$, $F_1 = b$ for different $(A, B, C, a, b) \in R$.

**Example 1**

Let $A = 3$, $B = 1.8$, $C = -1$, $F_0 = 1$, $F_1 = 5$. Then $F_{n+2} = 3 + 1.8F_{n+1} - F_n$, $n \geq 0$.

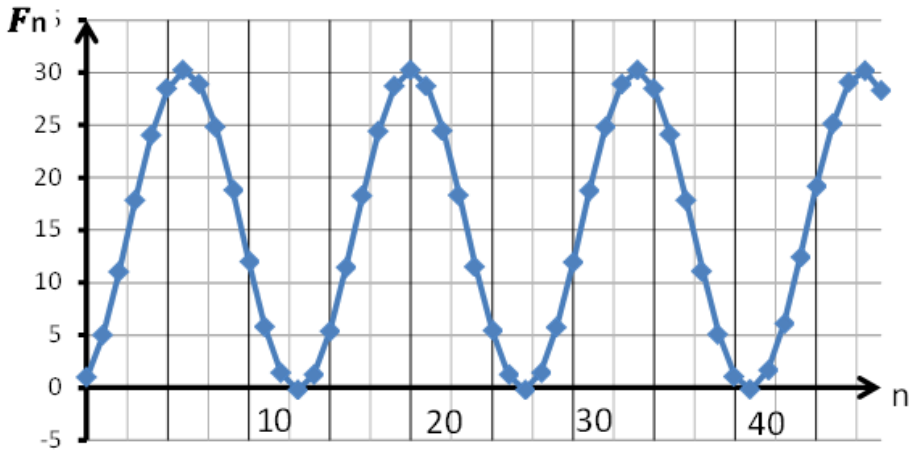

Figure 2. $F_n$ profile at n=0, 1,… 50. $F_{n+2} = 3 + 1.8F_{n+1} - F_n$, $n \geq 0$, $F_0 = 1$, $F_1 = 5$. $T = 13.93$.

For n=0,1,…14, the terms of $F_n$ are

| Fn= | 1.0 | 5.0 | 11.0 | 17.8 | 24.0 | 28.5 | 30.2 | 28.9 | 24.8 | 18.8 | 12.0 | 5.8 | 1.4 | -0.2 | 1.2 |
|---|---|---|---|---|---|---|---|---|---|---|---|---|---|---|---|
| n | 0 | 1 | 2 | 3 | 4 | 5 | 6 | 7 | 8 | 9 | 10 | 11 | 12 | 13 | 14 |

The profile of $F_n$ for n=1,…50 is shown in fig.2. $F_n$ is periodic with period $T = 13.93$, as it follows from formulas $T = \frac{2\pi}{\omega}$, $\omega = acos\frac{B}{2}$.

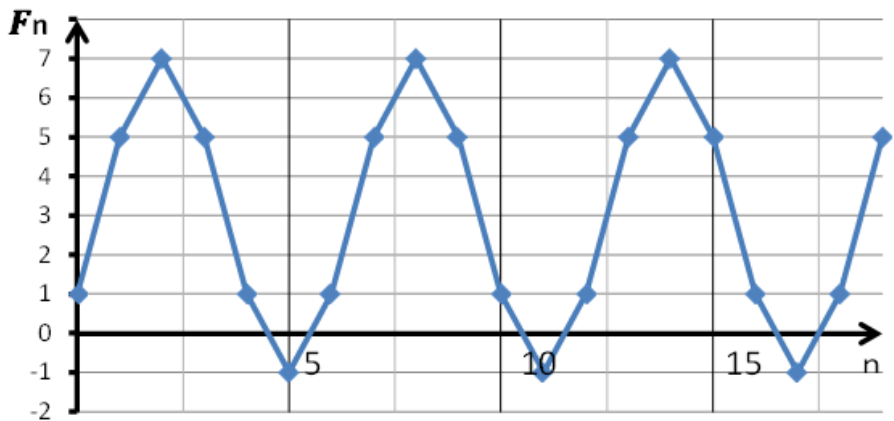

Figure 3. $F_n$ profile at n=0, 1,… 20. $F_{n+2} = 3 + F_{n+1} - F_n$, $n \geq 0$, $F_0 = 1$, $F_1 = 5$. $T = 6$.

**Example 2**

Let $A = 3$, $B = 1$, $C = -1$, $F_0 = 1$, $F_1 = 5$. Then $F_{n+2} = 3 + F_{n+1} - F_n$, $n \geq 0$. For n=0,1,…14, the terms of $F_n$ are

| Fn= | 1.0 | 5.0 | 7.0 | 5.0 | 1.0 | -1.0 | 1.0 | 5.0 | 7.0 | 5.0 | 1.0 | -1.0 | 1.0 | 5.0 | 7.0 |
|---|---|---|---|---|---|---|---|---|---|---|---|---|---|---|---|
| n | 0 | 1 | 2 | 3 | 4 | 5 | 6 | 7 | 8 | 9 | 10 | 11 | 12 | 13 | 14 |

Fig. 3 shows the profile of $F_n$ for n=1,...20. $F_n$ is periodic with period $T = 6$, as it follows from formulas $T = \frac{2\pi}{\omega}$, $\omega = acos\frac{B}{2}$.

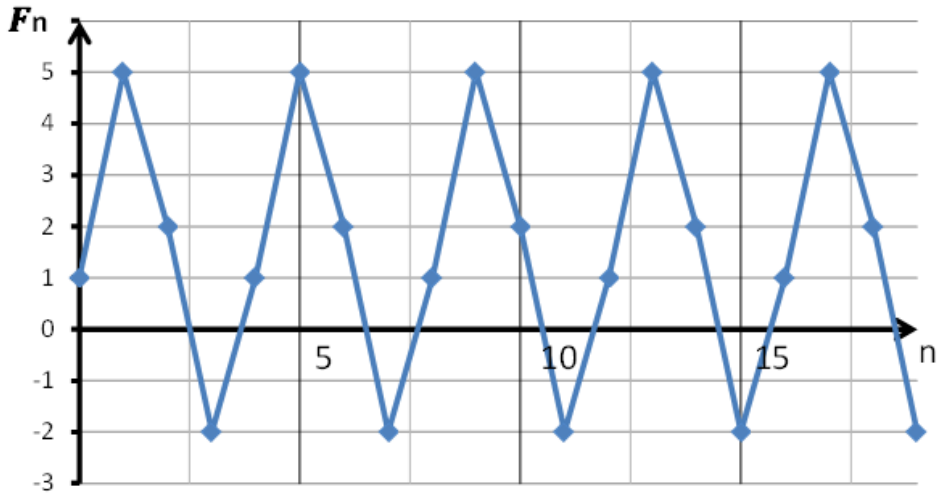

Figure 4. $F_n$ profile at n=0, 1,... 20. $F_{n+2} = 3 - F_n$, $n \geq 0$, $F_0 = 1$, $F_1 = 5$. $T = 4$.

**Example 3**
Let $A = 3$, $B = 0$, $C = -1$, $F_0 = 1$, $F_1 = 5$. Then $F_{n+2} = 3 - F_n$, $n \geq 0$. For n=0,1,...14, the terms of $F_n$ are

| Fn= | 1.0 | 5.0 | 2.0 | -2.0 | 1.0 | 5.0 | 2.0 | -2.0 | 1.0 | 5.0 | 2.0 | -2.0 | 1.0 | 5.0 | 2.0 |
|---|---|---|---|---|---|---|---|---|---|---|---|---|---|---|---|
| n | 0 | 1 | 2 | 3 | 4 | 5 | 6 | 7 | 8 | 9 | 10 | 11 | 12 | 13 | 14 |

Fig. 4 shows the profile of $F_n$ for n=1,...20. $F_n$ is periodic with period $T = 4$, as it follows from formulas $T = \frac{2\pi}{\omega}$, $\omega = acos\frac{B}{2}$.

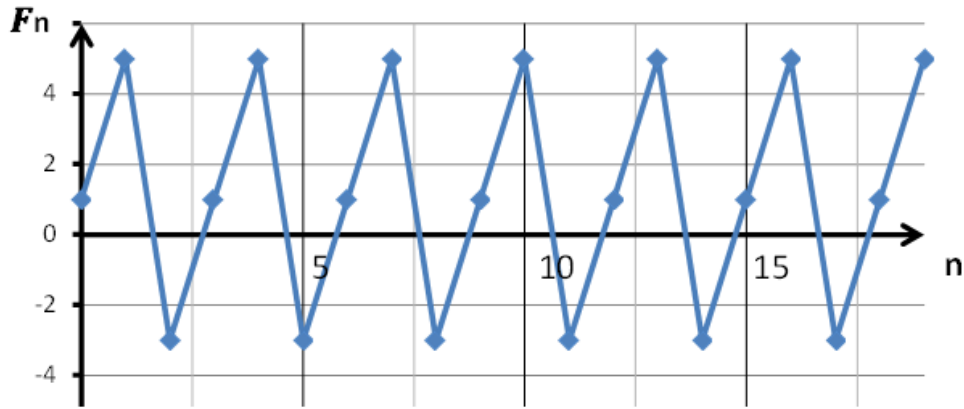

Figure 5. $F_n$ profile at n=0, 1,... 20. $F_{n+2} = 3 - F_{n+1} - F_n$, $n \geq 0$. $F_0 = 1$, $F_1 = 5$. $T = 3$.

**Example 4**
Let $A = 3$, $B = -1$, $C = -1$, $F_0 = 1$, $F_1 = 5$. Then $F_{n+2} = 3 - F_{n+1} - F_n$, $n \geq 0$. For n=0,1,...14, the sequence $F_n$ is

| Fn= | 1.0 | 5.0 | -3.0 | 1.0 | 5.0 | -3.0 | 1.0 | 5.0 | -3.0 | 1.0 | 5.0 | -3.0 | 1.0 | 5.0 | -3.0 |
|---|---|---|---|---|---|---|---|---|---|---|---|---|---|---|---|
| n | 0 | 1 | 2 | 3 | 4 | 5 | 6 | 7 | 8 | 9 | 10 | 11 | 12 | 13 | 14 |

The profile of $F_n$ for n=1,...20 is shown in fig. 5. $F_n$ is periodic with period $T = 3$, as it follows from formulas $T = \frac{2\pi}{\omega}$, $\omega = acos\frac{B}{2}$.

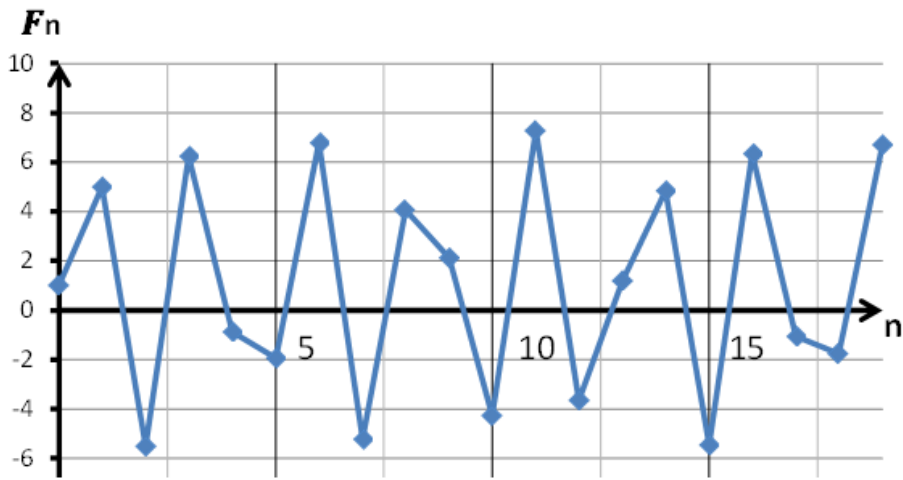

Figure 6. The profile of $F_n$ at n=0, 1,... 20. $F_{n+2} = 3 - 1.5F_{n+1} - F_n$, $n \geq 0$, $F_0 = 1$, $F_1 = 5$. $T = 2.6$.

**Example 5**
Let $A = 3$, $B = -1.5$, $C = -1$, $F_0 = 1$, $F_1 = 5$. Then $F_{n+2} = 3 - 1.5F_{n+1} - F_n$, $n \geq 0$. For n=0,1,...14, the sequence $F_n$ is

| Fn= | 1.0 | 5.0 | -5.5 | 6.3 | -0.9 | -1.9 | 6.8 | -5.2 | 4.1 | 2.1 | -4.3 | 7.3 | -3.6 | 1.2 | 4.9 |
|---|---|---|---|---|---|---|---|---|---|---|---|---|---|---|---|
| n | 0 | 1 | 2 | 3 | 4 | 5 | 6 | 7 | 8 | 9 | 10 | 11 | 12 | 13 | 14 |

The profile of $F_n$ for n=1,...20 is shown in fig. 6. $F_n$ is periodic with period $T = 2.51$, as it follows from formulas $T = \frac{2\pi}{\omega}$, $\omega = acos\frac{B}{2}$.

**6. Conclusion**
In connection with computational methods for approximating the solutions of partial differential equations, the generalized Fibonacci-like sequence $F_{n+2} = A + BF_{n+1} + CF_n$ with $F_0 = a$, $F_1 = b$, $n \geq 0$, where A, B, C, a and b are real numbers is investigated. It is shown that if $|B| < 2$, $C = -1$ then $\{F_n\}$ is periodic with the period $T > 2$. If period $T = 1$, then $A = C = 0$, $B = 1$, $F_{n+1} = F_n$, $n \geq 0$, $F_0 = F_1 = a$. If period $T = 2$ and $C \neq 1$ then $B = C - 1$, $D = \frac{A}{1-C}$, $F_{n+1} = D - F_n$, $n \geq 0$, $F_0 = a$, $F_1 = D - a$.
Numerical examples of periodic Fibonacci-like sequences are presented.